\documentstyle[preprint,aps,prb,epsfig]{revtex}

\begin{document}
     \draft
\title{Superconducting Alloys with Weak and Strong Scattering: Anderson's Theorem and a Superconductor-Insulator Transition.}

\author{ R. Moradian, J.F. Annett, B.L. Gy\"{o}rffy }

\address{
H.H. Wills Physics Laboratory, University of
Bristol, \\
Royal Fort, Tyndall Avenue, Bristol BS8 1TL, UK.}
\author{G. Litak} 
\address{ Department of Mechanics,Technical University of Lublin \\ 
Nadbystrzycka 36,PL-20-618 Lublin, Poland.}

\date{\today}
\maketitle

\begin{abstract}
We have studied the effects of strong impurity scattering on disordered superconductors beyond the low impurity concentration limit. By applying the full CPA to a superconductiong A-B binary alloy, we calculated the fluctuations of the local order parameters $\Delta_{A}, \Delta_{B}$  and charge densities, $n_{A}, n_{B}$ for weak and strong on site disorder. We find that for narrow band alloy s-wave superconductors the conditions for Anderson's theorem are satisfied in general only for the case of particle-hole symmetry. In this case it is satisfied regardless whether we are in the weak or strong scattering regimes. Interestingly, we find that strong scattering leads to band splitting and in this regime for any band filling we have a critical concentration where a superconductor-insulator quantum phase transition occurs at $T=0$.    
\end{abstract}


\section{Introduction}
As is well known, s-wave superconductivity is possible even in highly disordered systems. Examples are superconducting intermetalic alloys $($such as $Au_{c}Si_{1-c}, Nb_{c}Si_{1-c}$\cite{{Finkel'stein:94}}, $Mo_{1-c}Rh_{c}$\cite{Matthias:56}$)$, heavily doped cubic perovskites $($such as $Ba_{1-c}K_{c}BiO_{3}$ with $x\approx 0.4$$)$\cite{Batlogg:84} and chevrel phases $($ such as $Cs_{0.3}MoS_{2}$$)$.\cite{Woolam:74} What lies beyond these initially surprising facts is Anderson's theorem\cite{Anderson:59} according to  which pairing of time-reversed states leads to a finite gap $2\Delta$ in the density of states. Namely, the one particle states involved in the pairing need not be eigenstates of any translation operators and hence both $\Delta$ and $T_{c}$ are only weakly influenced by the disorder.

The key assumptions required for Anderson's theorem are $(i)$ non- magnetic scattering only $($time reversal symmetry$)$ $(ii)$ the self-consistent order parameter $\Delta$ does not fluctuate from configuration to configuration. Previously Gy\"{o}rffy {\it et al.}\cite{Gyorffy:97} examined the conditions for $(ii)$ to hold, and found that spatial fluctuations in $\Delta$ could be neglected provided that coherence length

\begin{equation}    
\xi=\frac{\hbar v_{F}}{\pi\Delta}
\label{eq:coherent length}
\end{equation}
is much greater than the lattice spacing, $a$. Anderson's theorem also follows from the classic theory of Abrikosov and Gorkov.\cite{Abrikosov:58,Abrikosov:59} Their argument is based on perturbation theory and the proof requires that the real part of the self energy $\Sigma(E)$ varies slowly  near the Fermi energy $E_{F}$.\cite{Rickayzen:80,Ketterson:99} It turns out that this is equivalent to the condition $\xi>>a$ in Eq.~\ref{eq:coherent length}.

The purpose of this paper is to examine disordered s-wave superconductors for which the spatial fluctuations of $\Delta$ cannot be neglected. The results of an such inquiry will be important for narrow band or short coherence length superconductors where $\Delta\le{E_{F}}$ or $\xi\sim a$, such as some of those mentioned above. In the work of Gy\"{o}rffy {\it et al.}\cite{Gyorffy:97} the order parameter fluctuations due to impurities were studied using a perturbative technique. Zhitomirsky and Walker\cite{Zhitomirsky:98} also calculated corrections to $T_{c}$ beyond Anderson's theorem due to order parameter fluctuations evaluated perturbatively. In the present paper the goal is to treat the spatially varying $\Delta$ and charge density fully self-consistently within the coherent potential approximation (CPA), and hence allow for the case of arbitrarily strong $\Delta$ or charge fluctuations.

In this regime a number of interesting new issues arise. For example Ghosal {\it et al.}\cite{Ghosal:98} showed that strong disorder scattering leads to strong spatial variations in $\Delta$, with the formation of superconducting 'islands' where $\Delta$ is large and other regions where $\Delta$ is small. Moreover, they found that the spectral gap persists even when $\Delta$ is very small in large regions of the sample. By contrast it was argued by Opperman\cite{Opperman:85} and Ziegler\cite{Ziegler:88} that order parameter fluctuations lead to a finite density of states within the gap. Similarly, in the calculations of the non-self-consistent density of states by Annett and Goldenfeld\cite{Annett:92} an impurity {\em band tail} develops due to spatial fluctuations in $\Delta$, and eventually this leads to gap-less s-wave superconductivity. Obviously the fluctuations in $\Delta$ can arise either due to randomness in the single particle site energy, $\varepsilon_{i}$, at each atomic site, as in an alloy, or due to randomness in the attractive interaction potential, $U_{i}$, as considered by Litak and Gy\"{o}rffy.\cite{Litak:99} Here we consider only the former case of alloy type disorder. 

In this paper we calculate the spatially random, self-consistent order parameter 
and charge due to AB binary alloy type disorder. The on site energy is $\varepsilon_{A}$
 on a fraction c of lattice sites and  $\varepsilon_{B}$ on the fraction $1-c$. The Green functions are calculated using the coherent potential approximation$(CPA)$\cite{Litak:92,Litak:98,Martin:99,Moradian:00}. This has been shown to be exact in the limit of infinite dimensions\cite{Moradian:00} and to reproduce the results of the self-consistent Born approximation for weak scattering $(\varepsilon_{A}-\varepsilon_{B}\ll W,$ where W is the bandwith $)$ and the self-consistent T-matrix approximation in the limit of low impurity concentration  ($c\ll 1$). We calculate the self-consistent A or B site order parameter $(\Delta_{A}, \Delta_{B})$ and charge density $(n_{A}, n_{B})$. Interestingly, we find that particle-hole symmetry leads to $\Delta_{A}=\Delta_{B}$, and we show that Anderson's theorem applies exactly in this case, even though $\xi\sim a$ or $\Delta\sim W$. A similar result was obtained previously for weak ($|U|\ll W$) \cite{Litak:a98} and strong
 $(|U|\gg W)$\cite{Litak:96} interactions, but only for the weak disorder limit. By contrast the result here is exact for both weak and strong scattering. Evidently, this latter result implies that $\xi\gg a$ is not a necessary condition for Anderson's theorem. Furthermore  in the extreme disorder limit $(\varepsilon_{A}-\varepsilon_{B}> W$) we have a disorder induced band-splitting in the normal state. Remarkably, if the chemical potential lies in the band gap there is a superconductor to insulator quantum phase transition at $T=0$. If the chemical potential lies inside one of the split bands, the normal state is metalic but there are impurity states inside the superconducting gap.

Below, in section II, we describe the model, and our CPA formalism and report our numerical results for the cases of weak $(\varepsilon_{A}-\varepsilon_{B}<< W)$ and intermediate  $(\varepsilon_{A}-\varepsilon_{B}< W)$ scattering.
In section III we show that the gap fluctuation vanishes in the case of particle-hole symmetry, whilst in section IV we argue that within CPA Anderson's theorem is exact in this case. Section V contain our results for the strong scattering $(\varepsilon_{A}-\varepsilon_{B}> W)$ case where we find a superconductor-insulator transition.

\section{ The model and CPA for s-wave superconducting alloys.}
We use an attractive U single band Hubbard model defined by the Hamiltonian:

\begin{equation}
H=-\sum_{ij\sigma}t_{ij}{c^{\dagger}_{i\sigma}}{c_{j\sigma}} +
\frac{1}{2}\sum_{i\sigma}U_{i}\hat{n}_{i\sigma}\hat{n}_{i-\sigma}
+ \sum_{i\sigma} (\varepsilon_{i}-\mu_{bare}) \hat{n}_{i\sigma} .
\label{eq:s-waveHamiltonian}
\end{equation}
where $c^{\dagger}_{i\sigma}$, $c_{i\sigma}$ are, respectively, the creation and
annihilation operators of electrons with spin $\sigma$ on the
lattice site $i$,
$\hat{n}_{i\sigma}=c^{\dagger}_{i\sigma}c_{i\sigma} $ is the local occupation number operator, $\mu_{bare}$ is
the chemical potential, and $t_{ij}$ is the hopping integral from
site $i$ to site $j$. $U_{i}$ is the attractive pairing interaction at site
i. In all of the numerical calculations shown below the interaction potential is $U_{i}=-3.2t$. For other interaction strengths we found similar results. $\varepsilon_{i}$ is the site diagonal random disorder potential which takes on values $\varepsilon_{A}$ with probablity c and $\varepsilon_{B}$ with probablity $1-c$.

After applying the Hartree-Fock-Gorkov (HFG) approximation to
Eq.~\ref{eq:s-waveHamiltonian} our task becomes a study of the Gorkov equation: 
\begin{eqnarray} 
 \sum_{l}&\left(\begin{array}{cc}
t_{il}+(\imath\omega_{n}+\mu_{bare}-\varepsilon_{i}-U_{i}n_{i\downarrow})\delta_{il} &
\Delta_{i}\delta_{il} \\ \Delta^{*}_{i}\delta_{il} & - t_{li}+(\imath\omega_{n} -\mu_{bare}+\varepsilon_{i}+n_{i\uparrow}U_{i})\delta_{il}              
                   \end{array} 
                           \right) \nonumber  \\
&\times\hat{G}(l,j,\imath\omega_{n})
=\delta_{ij} \left(\begin{array}{cc}
                                      1 & 0 \\ 0 & 1
                                      \end{array}
                                       \right).\label{eq:s-Bog de eq}
  \end{eqnarray}
where $\hat{G}(i,j,\imath\omega_{n})$ is the Fourier transform, with respect to the complex-time variable $\tau$, of the Greens function $\hat{G}(i,j,\tau)=\frac{1}{\beta}\sum_{\omega_{n}}e^{\imath\omega_{n}\tau}\hat{G}(i,j,\imath\omega_{n})$.The self-consistency conditions for the local order parameter and charge density are
\begin{equation}
\Delta_{i}=\frac{U_{i}}{\beta}\sum_{\omega_{n}}e^{\imath\omega_{n}\eta}G_{12}(i,i,\imath\omega_{n})
\label{eq:local order parameter}
 \end{equation}
and 
\begin{equation}
n_{i}=\frac{1}{\beta}\sum_{\omega_{n}}e^{\imath\omega_{n}\eta}G_{11}(i,i,\imath\omega_{n})
\label{eq:s wave charge density}
 \end{equation}
where $\eta$ is a positive infinitesimal. The task at hand is to solve the above equations for each configuration $\{\varepsilon_{i}\}$ of the site energy and pairing interaction, to obtain the local order parameter and charge  $\{\Delta_{i},,n_{i}\}$ and hence to calculate the configurationally averaged Greens function $\hat{\bar{G}}(i,i;\imath\omega_{n})=\langle \hat{G}(i,i;\imath\omega_{n};\{\varepsilon_{i}\}) \rangle$.


 Whilst most of the salient features of disordered superconductors are well described by the  Abrikosov and Gorkov(AG)\cite{Abrikosov:58,Abrikosov:59} theory, recently the CPA has also been brought to bear on the problem.\cite{Litak:92,Litak:98,Martin:99,Moradian:00} Within the CPA one replaces the on site 
random potential with a site independent self-energy matrix 
\begin{equation}
\hat{\Sigma}(\imath\omega_{n})=\left(\begin{array}{cc}
\Sigma_{11}(\imath\omega_{n}) & \Sigma_{12}(\imath\omega_{n}) \\
\Sigma_{21}(\imath\omega_{n}) & \Sigma_{22}(\imath\omega_{n})
 \end{array} \right)
 \end{equation}
 which is determined by the condition that an A or B impurity, corresponding to $\varepsilon_{A}$ and $\varepsilon_{B}$ respectively does not scatter on the average. Thus the average Green function is given by

\begin{eqnarray} 
 \sum_{l}&\left(\begin{array}{cc}
t_{il}+(\imath\omega_{n}+\mu-\Sigma_{11}(\imath\omega_{n}))\delta_{il} &
-\Sigma_{12}(\imath\omega_{n})\delta_{il} \\ 
-\Sigma_{21}(\imath\omega_{n})\delta_{il} & - t_{li}+(\imath\omega_{n} -\mu-\Sigma_{22}(\imath\omega_{n}))\delta_{il}              
                   \end{array} 
                           \right) \nonumber  \\
&\times\hat{G}^{c}(l,j,\imath\omega_{n})
=\delta_{ij} \left(\begin{array}{cc}
                                      1 & 0 \\ 0 & 1
                                      \end{array}
                                       \right).
\label{eq:cpa s-Bog de eq}
  \end{eqnarray}

In the previous application of the CPA to the above model\cite{Litak:98,Martin:99,Lustfeld:73} the condition which determined $\hat{\Sigma}$ was  implemented under the assumption that the pairing potential $\Delta_{i}$ does not fluctuate with the site energies. Therefore the self-consistency conditions (Eqs.~\ref{eq:local order parameter},~\ref{eq:s wave charge density}) were satisfied only on average. However, this presumption is not a necessary part of the CPA and, as will be seen presently, is unduely restrictive. Here, we consider the more general case where on an A or B site, with site energy $\varepsilon_{A}$ or $\varepsilon_{B}$, the pairing potential is allowed to be $\Delta_{A}$ or $\Delta_{B}$ respectively and the two local gaps, $\Delta_{A}$ and $\Delta_{B}$, are determined by the condition that the corresponding local gap equations (Eq.~\ref{eq:local
order parameter}) are separately satisfied.
Thus if the  probability that a site is occupied by an A atom is c and that for a B atom is $1-c$ the generalized CPA condition is
\begin{equation}
c\hat{T}_{A}(\imath\omega_{n})+(1-c)\hat{T}_{B}(\imath\omega_{n})=0
\label{eq:cpa condition}
 \end{equation}
where the single site T-matrices, $\hat{T}_{A}$ and $\hat{T}_{B}$, are given by
\begin{equation}
\hat{T}_{i}(\imath\omega_{n})=\hat{V}_{i}(\imath\omega_{n})\left(\hat{1}-\hat{G}^{c}(l,j,\imath\omega_{n})\hat{V}_{i}(\imath\omega_{n})\right)^{-1}
\end{equation}
in terms of the local, single site, scattering potential matrix
\begin{equation}
\hat{V}_{i}=\left(\begin{array}{cc}
            \varepsilon_{i}-\Sigma_{11}-\mu_{i} & -\Delta_{i}
-\Sigma_{12}\\-\Delta^{*}_{i}-\Sigma_{21} &-\varepsilon_{i}- \Sigma_{22}+\mu_{i}
\end{array}\right)
\label{eq:flct onsite potential}
\end{equation}
for $i=A,B$.

Note that in the above expression not only $\varepsilon_{i}$ and $\Delta_{i}$ are allowed to fluctuate but the local $\mu_{i}$ also take on different values on A and B sites. Evidently, such variations arise from charge fluctuations. In the present HFG approximation this is described by
 
\begin{equation}
\mu_{i\sigma}=\mu_{bare}-\frac{1}{2}n_{i\sigma}U_{i}.
\label{eq:chemical-ptential}
 \end{equation}
Clearly, once the above CPA problem has been solved for a set of $\varepsilon_{A},\varepsilon_{B}, \Delta_{A}, \Delta_{B}, n_{A}$ and $n_{B}$ self-consistency requires that they are recalculated using the relations

\begin{equation}
\Delta_{A,B}=\frac{U_{A,B}}{\beta}\sum_{\omega_{n}}e^{\imath\omega_{n}\eta}G^{A,B}_{12}(i,i;\imath\omega_{n})
\end{equation}

\begin{equation}
n_{A,B}=\frac{1}{\beta}\sum_{\omega_{n}}e^{\imath\omega_{n}\eta}G^{A,B}_{11}(i,i;\imath\omega_{n})
\end{equation}
where $\hat{G}^{A,B}(i,i;\imath\omega_{n})$ is the Greens function matrix averaged over all configurations with an A or B atom, respectively, on the site i.
In our calculation we assume, the non-magnetic case, that
$n_{i\uparrow}=n_{i\downarrow}=\frac{1}{2}{n}_{i}$, and hence

\begin{equation}
\mu_{A,B}=\mu_{bare}-\frac{1}{2}U_{A,B} n_{A,B}.
\end{equation}
Moreover, the patially averaged Green functions $\hat{G}^{A,B}(i,i;\imath\omega_{n})$ are approximated by the Green function for A or B impurity in the CPA effective lattice described by the self-energy $\hat{\Sigma}$. They are given by

\begin{equation}
\hat{G}^{A,B}(i,i;\imath\omega_{n})=\hat{G}^{c}(i,i;\imath\omega_{n})+\hat{G}^{c}(i,i;\imath\omega_{n})\hat{T}_{A,B}\hat{G}^{c}(i,i;\imath\omega_{n})
\label{eq:CPA impurity Green function}.
\end{equation}

In short, equations $6-15$ fully specify a self-consistent procedure, which when carried to convergence, constitutes the complete CPA for disordered superconductors, in the HFG approximation, for the model Hamiltonian in Eq.~\ref{eq:s-waveHamiltonian}. Note that since the CPA is the mean field theory of disorder and the HFG approximation is that for superconductivity, the above theory should be regarded as the mean field theory which treats disorder and electron interaction simultaneously and on an equal footing. By treating the self-consistency only on average, earlier works\cite{Litak:98,Martin:99,Lustfeld:73} did not include the effects of fluctuations in $\Delta$ and $n$. Previously the full CPA was only implemented to determine the influence of $\Delta_{i}$ fluctuation on $T_{c}$.\cite{Litak:92} In the remainder of this paper we investigate the full consequences of treating interaction and disorder together on the basis of Eqs. $6-15$.

 Our results for the gap and charge fluctuations as functions of the average band filling $\bar{n}$ are shown in Figs.$1$ and $2$. 
To simplify matters we used a $2d$ square lattice with lattice constant $a=1$ and band energy

\begin{equation}
\epsilon_{\bf
k}=-2t(cosk_{x}+cosk_{y})   
\end{equation}
where t is the nearest neighbour hopping amplitude. In these calculations we have taken the energy difference $\delta=\varepsilon_{A}-\varepsilon_{B}$ to be a significant fraction of the bandwidth $W=8t$ $(\delta=0.5t , \delta=2t)$ and hence we are in the fairly strong scattering regime  $($ for $\delta=2t)$. As Fig.$1$ shows $\Delta_{A}\neq\Delta_{B}$ except at the point $\bar{n}=1$, namely a half filled band. Clearly from Fig.$2$ $n_{A}\neq n_{B}$ at any filling except $0$ or $2$. Thus, unlike in previous calculations\cite{Martin:99} the $\Delta$ and $n$ fluctuations are central features of our results. For emphasis we show in
 Fig.$3$ the standard deviations of the order parameter    

\begin{equation}
M_{\Delta}=\langle(|\delta\Delta_{i})^{2}|\rangle=\langle|\Delta^{2}_{i}|\rangle
-\langle |\Delta_{i}| \rangle^{2}\cong c(1-c)(\Delta_{A}-\Delta_{B})^{2}
\label{eq:order-para-moment} 
\end{equation}
and charge density
\begin{equation}
M_{n}=\langle(\delta n_{i})^{2}\rangle=\langle n^{2}_{i}\rangle
-\langle n_{i} \rangle^{2}\cong c(1-c)(n_{A}-n_{B})^{2}
\label{eq:charge densit-moment}
\end{equation}
as predicted by our CPA calculations.

Remarkably, at half filling the fluctuations in the pairing potential go to zero, whilst the charge density fluctuations are at their strongest. To investigate the origin of this interesting phenomena we studied the case where $c=0.75\ne 0.5$ but the band is still half filled at $\bar{n}=1$. For this case $\Delta_{A}, \Delta_{B}$ and $\bar{\Delta}$ are shown in Fig.$4$ as a function of temperature. Evidently for all $T<T_{c}$, $\Delta_{A}\ne\Delta_{B}$ and hence $M_{\Delta}\ne 0$. In what  follows we unravel the root cause of this behavior.

\section{particle-hole symmetry}

Recall that in Figs.$1-3$ the order parameter fluctuations vanish, $\Delta_{A}=\Delta_{B}$, for the case $\bar{n}=1$ and $c=0.5$, namely equal concentrations of A and B atoms. This special case is one where the Fermi energy is at the center of the band and the density of states is symmetric, and hence  particle-hole symmetry occurs. For this case and only this case $\mu=0$, $\mu_{bare}=\frac{1}{2}U$ and $\mu_{B}=-\mu_{A}$. Furthermore, since the self-energy always obey the same symmetries as the Green functions  

\begin{eqnarray}
  \Sigma_{11}(\imath\omega_{n})=- \Sigma^{*}_{22}(\imath\omega_{n})\nonumber\\
  \Sigma_{12}(\imath\omega_{n})=\Sigma^{*}_{21}(\imath\omega_{n}),
\label{eq:sigma symmetry}
\end{eqnarray}
for particle-hole symmetry there is also the property that

\begin{equation}
 \Sigma_{11}(\imath\omega_{n})=- \Sigma^{*}_{11}(\imath\omega_{n}).
\label{eq:sig-p-h}
\end{equation}
Namely,

\begin{equation}
 \Re\Sigma_{11}(\imath\omega_{n})=0
\label{eq:relsigma}
\end{equation}
and consequently
\begin{equation}
\Re G^{c}_{11}(\imath\omega_{n})=0.
\label{eq:realgreen}
\end{equation}

Noting that the CPA respects these symmetries we can rewrite the CPA
condition in Eq.~\ref{eq:cpa condition}, as 
\begin{equation}
c{\hat{V}_{B}}^{-1}+(1-c){\hat{V}_{A}}^{-1}=\hat{G}^{c}(i,i;\imath\omega_{n}).
\label{eq:2cpa-condtion}
\end{equation}
Combining this with particle-hole symmetry as described by Eqs.~\ref{eq:sigma symmetry},~\ref{eq:sig-p-h},~\ref{eq:relsigma},~\ref{eq:realgreen} implies that 

\begin{equation}
(\Delta_{A}-\Delta_{B})\left(\frac{\delta}{2}\Sigma_{12}+\frac{\delta}{2}\Sigma_{21}+(\mu_{A}-\frac{\delta}{2})\Delta_{A}+(\mu_{A}-\frac{\delta}{2})\Delta_{B}\right)=0.
\end{equation}
and hence, since the second bracket is non-zero, that
 
\begin{equation}
\Delta_{A}=\Delta_{B}.
\end{equation}
Therefore particle-hole symmetry implies the absence of the order parameter fluctuations.

\section{Proof of Anderson's theorem in the particle-hole symmetric density of state case}

In non-magnetic disordered local s-wave superconductors the traditional argument leading to Anderson's theorem assumes that the fluctuations of the order parameter are negligible, $\Delta_{i}\approx \bar{\Delta}$. Anderson's theorem shows  that in this case there are no bound states between quasi-particles  and impurity sites and therefore the quasi-particle energy gap is absolute. Namely, there are no impurity states inside of the gap.\cite{Anderson:59} Furthermore $T_{c}$ is found simply by replacing, the clean system density of states $N(E)$, with its disordered system average $\bar{N}(E)$ in the gap equation.

In an alternative route to the same result Abrikosov and
 Gorkov\cite{Abrikosov:58,Abrikosov:59} use perturbation theory which implies that the self-energy can be approximated by

\begin{equation}
\Sigma_{11}(\imath\omega_{n})=-\frac{\imath}{\tau}sign(\omega_{n})
\end{equation}
where $\tau$ is a wave vector and frequency independent quasi-particle life time. This is justified in the case of the non-self-consistent Born approximation by the assumption that the relevant energy scale is $|\omega_{n}|\leq\omega_{D}$ $(\omega_{D}$ is Debeye frequncy$)$ and hence only states near the Fermi surface are relevant. More implicitly it is assumed that  near the Fermi surface the density of states is a constant and hence there is effectively particle-hole symmetry.

However in general we cannot assume that $\Sigma_{11}(\imath\omega_{n})$ is of this form. In the case of narrow band superconductors, there is no Debeye cut off and so one cannot assume that only states near the Fermi level are significant. In particular $\Im\Sigma_{11}(\imath\omega_{n})$ will not be a constant, and $\Re\Sigma_{11}(\imath\omega_{n})$ need not be zero. Nevertheless, if for some reason particle-hole symmetry is obeyed, Anderson's theorem will obtain in full CPA, self-consistent Born, and T-matrix approximations.

In Figs.$5$ and $6$ we illustrate using our explicit calculation the energy dependence of $\Re\Sigma_{11}(E)$, $\Im\Sigma_{11}(E)$, $\Re(R(E))$ and $\Im(R(E))$ for the cases of particle-hole symetric or nonsymmetric cases. Note that function $R(E)$ is defined by 

\begin{equation}
R(E)=\frac{1}{2}\left (\Sigma_{11}(E+\imath 0^{+})+\Sigma_{11}(-E-\imath 0^{+})\right),
\end{equation}
and $R(E)$ is the analytical continuation of $\Re\Sigma_{11}(\imath\omega_{n})$ to the real axis. Figs.$5$ and $6$ show that for particle-hole symmetry, $\bar{n}=1$ and $c=0.5$, $R(E)$ is equal to zero but in other cases  $R(E)\neq 0$.
We shall now analyze the consequences of the particle-hole symmetry $E\rightarrow -E$. 

The Green function for CPA, self-consistent Born or T-matrix approximations can be written in the form 
\begin{equation}
\hat{\bar{G}}(i,i;\imath\omega_{n}) = \frac{1}{N}\sum_{k}{ 
                        \left(\begin{array}{cc}
                   \imath{\tilde{\omega}}_{n}-\epsilon_{k}+\tilde{\mu}
 & \tilde{\Delta} \\ {\tilde{\Delta}}^\ast 
&  \imath{\tilde{\omega}}_{n}+\epsilon_{k}-\tilde{\mu}
\end{array}\right)}^{-1}
\end{equation}
where the renormalized parameters $\tilde{\Delta}$, $\tilde{\omega}_n$
 and
$\tilde{\mu}$ and $\eta_\omega$ are given by
\begin{eqnarray} 
 {\tilde{\omega}}_{n} & = & \omega_{n}(1-\frac{
\Im{\Sigma}_{11}(\imath \omega_n)
}{\omega_{n}})
 \nonumber \\ 
 \tilde{\Delta} & = & \Delta (1-\frac{\Im{\Sigma}_{11}(\imath \omega_n)}{\omega_{n}})    
 \nonumber \\ 
\tilde{\mu} & = & \mu-\Re\Sigma_{11}(\imath \omega_n)    
 \nonumber \\ 
  \eta_{\omega} & = & (1-\frac{\Im{\Sigma}_{11}(\imath \omega_n)}{\omega_{n}}).
\end{eqnarray}
Note that these  renormalized parameters are the same as in the original
paper of Abrikosov and Gorkov\cite{Abrikosov:59}, where
$\hat{\Sigma}$ was computed in the non-self-consistent Born approximation. They found that $\Re\Sigma_{11}$ 
 near the Fermi surface is independents of $\omega$, and therefore eventually it can be absorbed into the chemical
potential. The same procedure was followed  Martin et al..\cite{Martin:99}

However we proceed without these simplifications and retain the full energy dependence of   $\Re\Sigma_{11}(E)$ and $\Im\Sigma_{11}(E)$.  
Now, particle-hole symmetry implies that  $\Re\Sigma_{11}(\imath\omega_{n})=0$, and therefore $\tilde{\mu}=\mu$. Using the above relations the gap equation becomes 

\begin{equation}
1=\frac{|U|}{\beta}\sum_{n}
\int^{\infty}_{-\infty}
\frac{N(\tilde{\epsilon},\imath \omega_n)}{\omega^2_{n}+
\tilde{\epsilon}^2 + |\Delta|^2}
d\tilde{\epsilon}\label{eq:Anderson}
\end{equation}
where
\begin{equation}
N(\tilde{\epsilon},\imath \omega_n)=\frac{1}{N}\sum_{k}\frac{1}{\eta_{\omega}}
\delta(\tilde{\epsilon}-{\tilde{\epsilon_k}})
\end{equation}
and, $\tilde{\epsilon_k}=\frac{\epsilon_{k}}{\eta_{\omega}}$ is the
renormalized band energy.

Surprisingly in the case of particle-hole symmetry the quantity $N(\tilde{\epsilon},\imath \omega_n)$
in Eq.~\ref{eq:Anderson} becomes equal to the disorder
average normal state density of states, namely, $\bar{N}(\tilde{\epsilon})$ and hence, without further assumptions, Anderson's theorem obtains. To be quick and explicit we note that this last step follows from
the property of a delta function that
$\delta(\frac{x}{a})=a\delta(x)$, and hence 
\begin{equation}
N(\tilde{\epsilon},\imath \omega_n )=\frac{1}{N}\sum_{k}
\frac{1}{\eta_{\omega}}\delta(\tilde{\epsilon}-
\tilde{\epsilon_k})=N(\eta_{\omega}\tilde{\epsilon})=\bar{N}(\tilde{\epsilon}),
\label{eq:average density of state}
\end{equation}
In short, we again have Anderson's theorem that $T_c$ is given by the
usual gap equation, but with the disorder average normal density of states. However, unlike other proofs,ours does not neglect the energy dependence of $\Re\Sigma_{11}(E)$ or $\Im\Sigma_{11}(E)$.

\section{Superconductors in the split band regime}

One of the main virtues of the CPA in the normal state is the fact that it describes band splitting correctly.\cite{Eliott:74} Namely, for $\varepsilon_{A}-\varepsilon_{B}$ less than the bandwidth, $8t$ in our case, it predicts an effective band somewhere in between the bands of pure A or pure B metal, whilst for $\varepsilon_{A}-\varepsilon_{B}$ bigger than the half bandwidth CPA predicts two, smeared, but nevertheless well defined, bands seperated by a gap. The two bands are at energies where there would have been an A or B band in one of the pure systems. In this later case the wave function corresponding to the A band is large mainly on the A sites and that in the B band is significant only on the B sites. In what follows we shall investigate superconductivity in this split band regime of CPA.

To investigate the consequences for superconductivity in the above split band regime we have solved Eq.6-15 using a strong scattering potential $(\varepsilon_{A}-\varepsilon_{B}\sim W)$. From these solutions two interesting points emerged. Firstly we found that for the particle-hole symmetric case, reported in Fig.$8$, $\Delta_{A}=\Delta_{B}$ and hence even in this strong scattering state there are no fluctuations in $\Delta$. Nevertheless, scattering has a large effect on the superconductivity through the configurationally average density of states $\bar{N}(E)$ in the Eq.~\ref{eq:average density of state}. Namely as shown in Fig.$8$ $\Delta\rightarrow 0$ for the critical strength of scattering $\delta\simeq 4.5t$. 

The other interesting phenomenon is a superconductor to insulator transition. Recently Scalettar, Trivedi and Huscroft\cite{Scalettar:98} discovered a superconductor-insulator transition in the disordered attractive U Hubbard model, using Monte Corlo simulations. They found that for strong disorder the superconducting gap is replaced by an insulating gap, both for weak and strong interaction U. Clearly the superconductor to insulator transition in our CPA calculaton has a similar origin, although our model of disorder is different. 

In general the band filling $\bar{n}$ is given by 

\begin{equation}
\bar{n}=(1-c)n_{B}+cn_{A}.
\label{eq:filling}
\end{equation} 
where $n_{A}$ and $n_{B}$ are the partial averaged occupation numbers on A or B sites respectively. We shall discuss three different cases of Eq.~\ref{eq:filling}: case $(i)$ $(n_{A}=0, n_{B}<2)$, case $(ii)$ $(n_{A}=0, n_{B}=2)$ and case  $(iii)$ $(n_{A}\neq 0,n_{B}=2)$. In the first case the 
 A band is empty and the filling of the B band$(n_{B})$ is

\begin{equation}
n_{B}=\frac{\bar{n}}{1-c}.
\label{eq:bfilling}
\end{equation} 
For the second case the B band is completely, doubly, occupied and the A band completly empty.In the third case the A band is partially occupied with 

\begin{equation}
n_{A}=\frac{\bar{n}-2(1-c)}{c}.
\label{eq:afilling}
\end{equation} 
and the B band is fully occupied. The second case is a special case of Eq.~\ref{eq:afilling}. Fig.$7$ shows the A, B and average  normal density of states for these three different concentrations. In Fig.$7(a)$ the A sites are approximately empty $(n_{A}\approx 0)$ but there is more than one electron on the B sites. The graph plotted was for the case $\bar{n}=1$ and $c=0.25$, and therefore from Eq.~\ref{eq:bfilling} we have $ n_{B}= \frac{4}{3}$. Fig.$7(a)$ is a band metal with hopping of electrons from the B sites to the B sites.
 In Fig.$7(b)$ the A sites are almost completely empty, $n_{A}= 0$, and the B sites are doubly occupied, $n_{B}= 2$. Therefore there is a gap at the Fermi energy and the system is a band insulator. In Fig.$7(c)$ the B band is fully occupied while the A band is partially filled. For this case we had $\bar{n}=1$ and $c=0.75$ and so by Eq.~\ref{eq:afilling}, the band filling of the A sites is $n_{A}=\frac{2}{3}$. Similarly Fig.$7(c)$ is a metal band with hopping from the A sites to the A sites. In Fig.$7(b)$ there is no hopping and we can regard this state as a band insulator.

As illustrated in Fig.$8a$, in the split band regime the superconductor gap closes,$\bar{\Delta}\rightarrow 0$ as $\delta\rightarrow 4.5t$ but is replaced by an insulating gap for $\delta > 4.5t$ . Evidently, this can also happen in the non particle-hole symmetric case as shown in Fig.$8b$. The general condition for the superconductor-insulator transition is saturation of the B sites with $2$ electrons of opposite spin while the A sites are empty. This happens when 
 
\begin{equation}
\bar{n}=2(1-c).
\label{eq:transition}  
\end{equation}
In this case the Fermi level lies in the gap and all the B states are filled and the A states are empty. For completeness we show the suppression of $\bar{\Delta}$ by disorder for particle-hole symmetric and non particle-hole symmetric cases in Fig.$9$, obeying the condition from Eq.~\ref{eq:transition}.

Now numerically we shall test our prediction, in Eqs.~\ref{eq:afilling},~\ref{eq:bfilling}, for the two non-particle-hole symmetric cases:$c=0.25, \bar{n}=1$ and $c=0.75, \bar{n}=1$, as discussed at the beginning of this section. In this cases there is no superconductor-insulator transition. For a greater understanding of the details of the band splitting  mechanism we plot the density of states, the order parameter and the charge density on A and B sites. In the first case the concentration of A sites is less than B sites, but in the second case the concentration of A sites is more than of B sites. As expected, in Fig.$10(a)$ we see that there are two split bands, a normal empty band on the A sites and a superconducting band on B sites. Conversely in Fig.$10(b)$ the B band is a normal doubly occupied band and the A band is the superconducting band.

The effect of disorder on the average order parameter and the A and B sites local order parameters $(\bar{\Delta}, \Delta_{A}, \Delta_{B})$ and charge densities $(n_{A},n_{B})$ are shown in  Figs.$11$ and Fig.$12$. One can see that, for weak scattering the local order parameters and charge densities of A and B sites are approximatly the same but with increasing disorder the difference between them will increase, with one of $\Delta_{A}$ or $\Delta_{B}$ going to zero$($depending on the relative concentrations$)$. In Fig.$11(a)$ and Fig.$12(a)$ the concentration of A sites is less than that of B sites therefore $\Delta_{A}\rightarrow 0$, $\Delta_{B}\rightarrow constant$ while $n_{A}\rightarrow 0$, $n_{B}\rightarrow \frac{4}{3}$. In contrast to this, in Fig.$11(b)$ and Fig.$12(b)$ the concentration of A sites is more than that of B sites, consequently $\Delta_{A}\rightarrow constant$, $\Delta_{B}\rightarrow 0$ and $n_{A}\rightarrow \frac{2}{3}$, $n_{B}\rightarrow 2$.

Schematically Fig.$13$ shows the physical mechanism of the band splitting in terms of A and B lattice sites. Clearly when all B sites become doubley occupied and all the A sites become empty there is no hopping and the system becomes insulating. Therefore as a function of the disorder, $\delta$, or band filling $\bar{n}$ there is a $T=0$ superconductor-insulator quantum phase transition. This transition corroborates the suggestion of Scalettar, Trivedi and Huscroft\cite{Scalettar:98} that disorder in superconductors can lead to a superconductor-insulator transition. However it shuld be noted that they used a uniform random distribution of site energies, $\varepsilon_{i}$, wheras we used a binary alloy model.

\section{Conclusion}

In this paper, for the first time, we applied the full CPA for disorder superconductors. In our calculations the self-consistency equations were solved fully within CPA so that self-consitency was properly satisfied on each type of CPA impurity sites, and not just on the average. It is in this form that the CPA-HFG is the only `controled' mean field theory of disordered superconductors.\cite{Moradian:00}

We plotted the local order parameter and charge density of A or B atoms of a binary alloy for both weak and strong scattering limits. We found that only for one  special case are the order parameters of A and B sites equal, and consequently the fluctuations of $\Delta$ are zero. For this point we have shown analytically that the condition for Anderson's theorem is fullfilled not only for weak scattering but also for strong scattering in a particle-hole symmetric band. By contrast neither, the density of states nor $T_{c}$ is exactly constant, although in the gap equation, the quasi-particle energy gap is absolute. In the gap equation the  normal clean system density of states in the gap equation is replaced by the normal disorder average density of states, and therefore the only changes of the superconducting density of states and $T_{c}$ come from the latter.

In narrow band binary alloy s-wave superconductors, we showed that strong disorder leads to two different interesting phenomena: $(i)$ band splitting with a quantum superconductor-insulator phase transition at $T=0$, $(ii)$ band splitting without a phase transition. In this last case in terms of concentration and average band filling, one band is normal $($doubly occupied or empty$)$ and superconductivity is only present in the another band that is partially  occupied.

\acknowledgements
This work has been supported by Ministry of Science, Reaserch and Technology of Iran under grant number GR/752066.

\vfill
\begin{figure}
\centerline{\epsfig{file=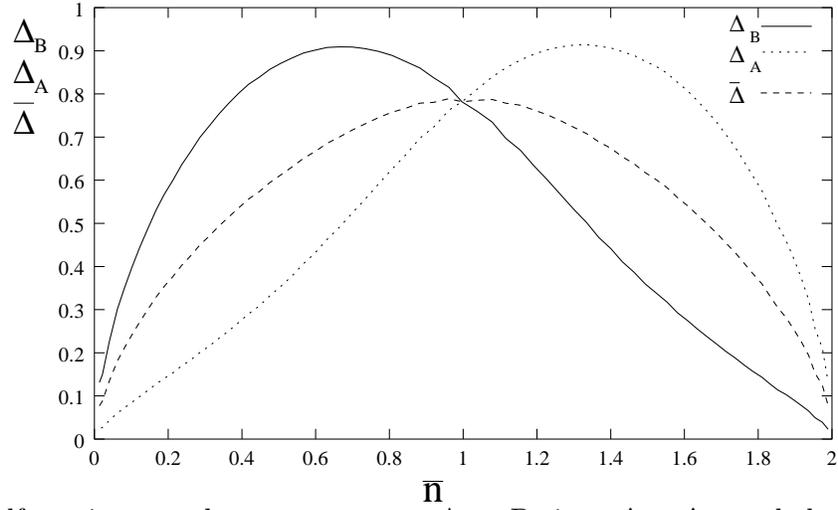,width=11.0cm,angle=0}}
\caption{  Self-consistent order parameters at A or B sites, $\Delta_{A}$, $\Delta_{B}$ and the average
$\bar{\Delta}$ as a function of band filling. Here $T=0.008625t$,$c=0.5$ and $\delta=2t$. Note that at half band filling
$\Delta_{A}=\Delta_{B}=\bar{\Delta}$ showing the fluctuation vanishes when
there is particle-hole symmetry.
\label{figure1}}
\end{figure}.

\vfill
\begin{figure}
\centerline{\epsfig{file=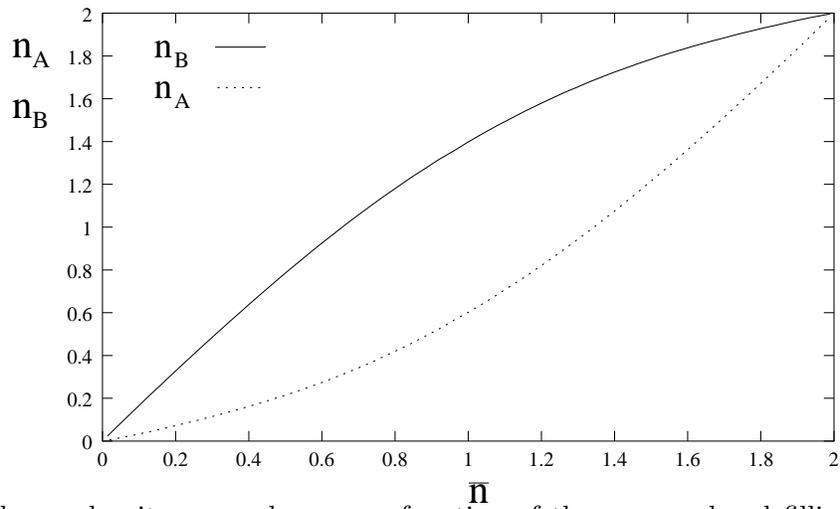,width=11.0cm,angle=0}}
\caption{ Charge density $n_{A}$ and $n_{B}$ as a
function of the average band filling $\bar{n}$ for $c=0.5$, $\delta=2t$ and
temperature is $T=0.008625t$. Note that the $n_{A}$, $n_{B}$ are never equal, except at $\bar{n}=0$ or $2$.
\label{figure2}}
\end{figure}.

\vfill
\begin{figure}
\centerline{\epsfig{file=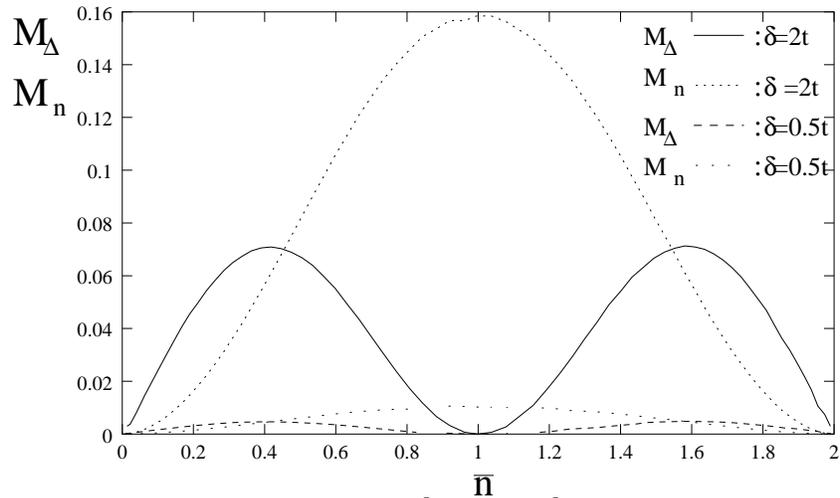,width=11.0cm,angle=0}}
\caption{ Fluctuation in $\Delta$, $M_{\Delta}=\langle
|\Delta^{2}_{i}|\rangle-{|\Delta^{c}}|^{2}$, and fluctuation of charge density,
 $M_{n}=\langle| n^{2}_{i}|\rangle- |{n^{c}}|^{2}$, as a function of band
filling for $c=0.5$ and $T=0.008625t$. Note that the $\Delta$ fluctuation is zero at half band
filling, while the charge density fluctuation is maximum. Therefore $\bar{n}=1$
obeys the conditions for Anderson's theorem: particle-hole symmetry and as a consequence of this there is absence of fluctuation in $\Delta$. For other fillings this condition is not true. 
\label{figure3}}
\end{figure}

\vfill
\begin{figure}
\centerline{\epsfig{file=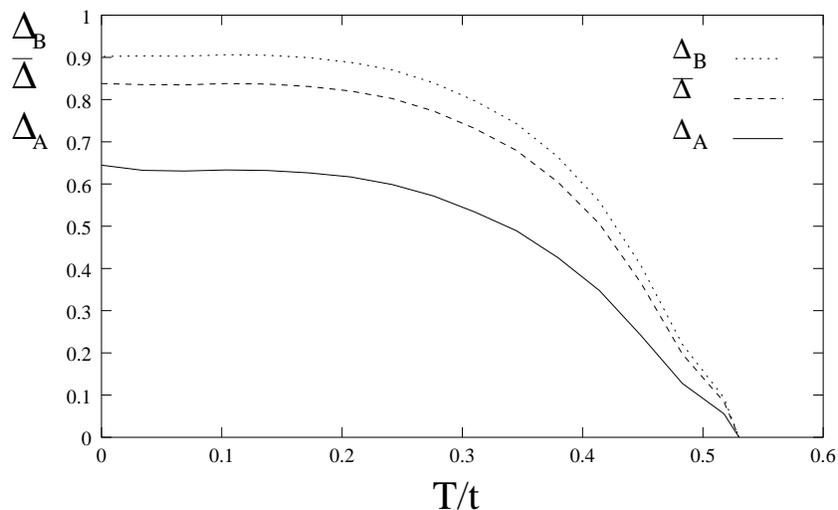,width=11.0cm,angle=0}}
\caption{ $\Delta_{A}$, $\Delta_{B}$ and the average $\bar{\Delta}$ as a function of temperature in the  particle-hole asymmetric
case for $\bar{n}=1$, $c=0.75$ and $\delta=2t$. Note that in this case $\Delta_{A}$, $\Delta_{B}$ and the average
$\bar{\Delta}$ go to zero at $T_{c} \simeq 0.525t$.
\label{figure4}}
\end{figure}.

\vfill
\begin{figure}
\centerline{\epsfig{file=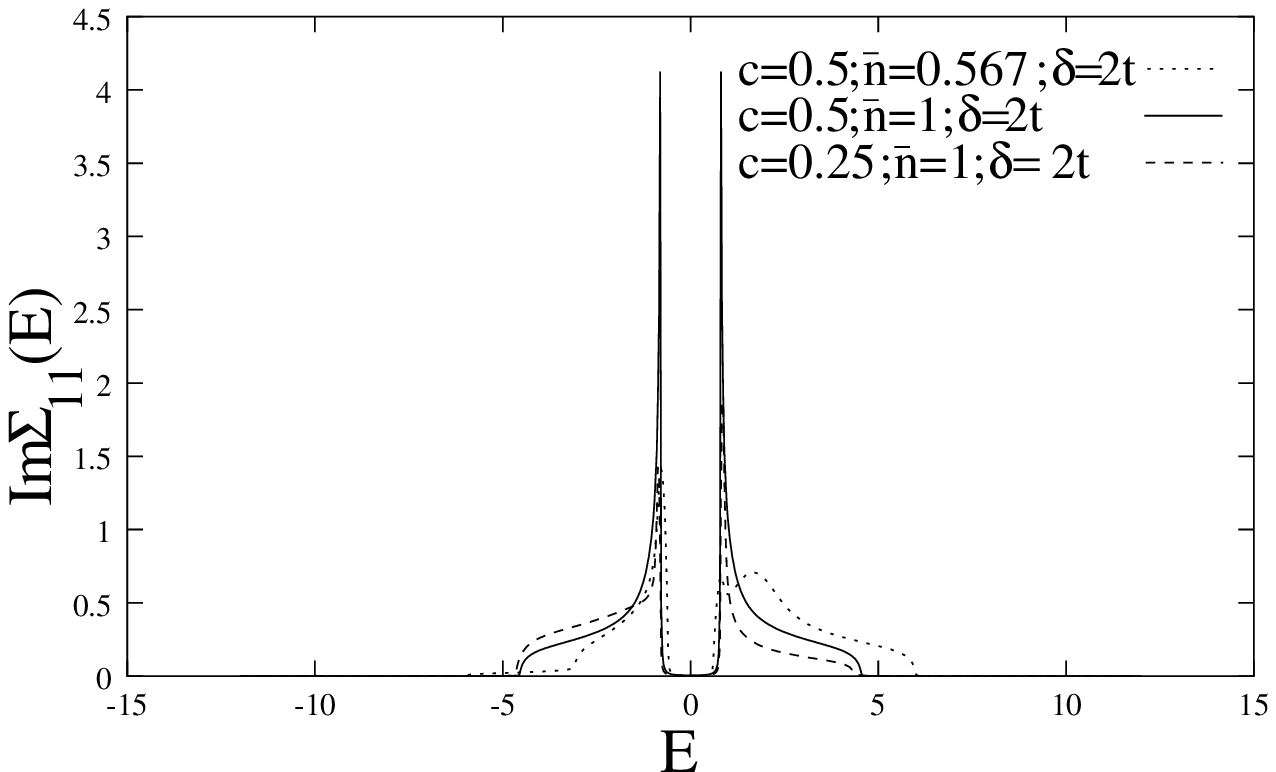,width=11.0cm,angle=0}}
\centerline{\epsfig{file=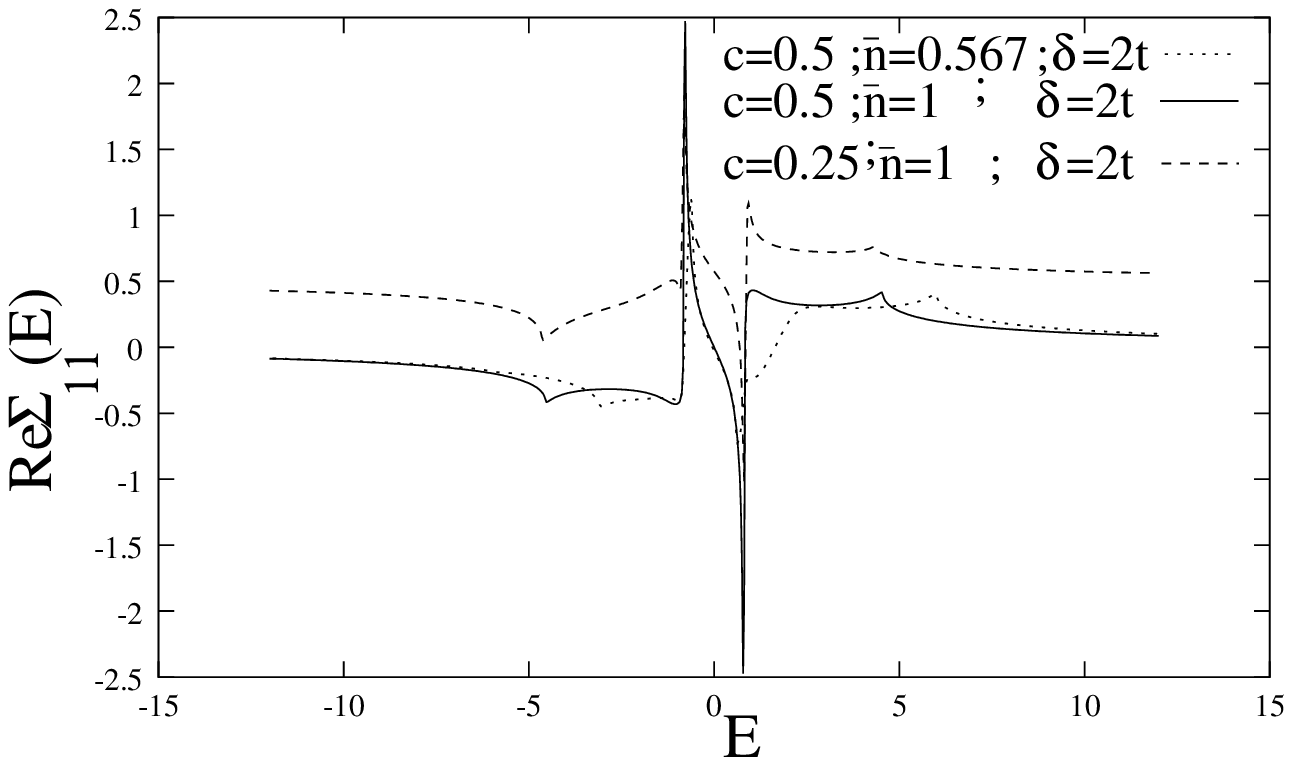,width=11.0cm,angle=0}}
\caption{$\Re\Sigma_{11}(E)$ and $\Im\Sigma_{11}(E)$ as a function of energy E. The case $c=0.5, \bar{n}=1$ has a  particle-hole symmetry and others are non-particle-hole symmetric. As it is obvious that for such a narrow band superconductors neither $ \Re\Sigma_{11}(E)$ nor $\Im\Sigma_{11}(E)$ are constant, unlike the weak scattering Born approximation limit.
\label{figure5}}
\end{figure}.

\vfill
\begin{figure}
\centerline{\epsfig{file=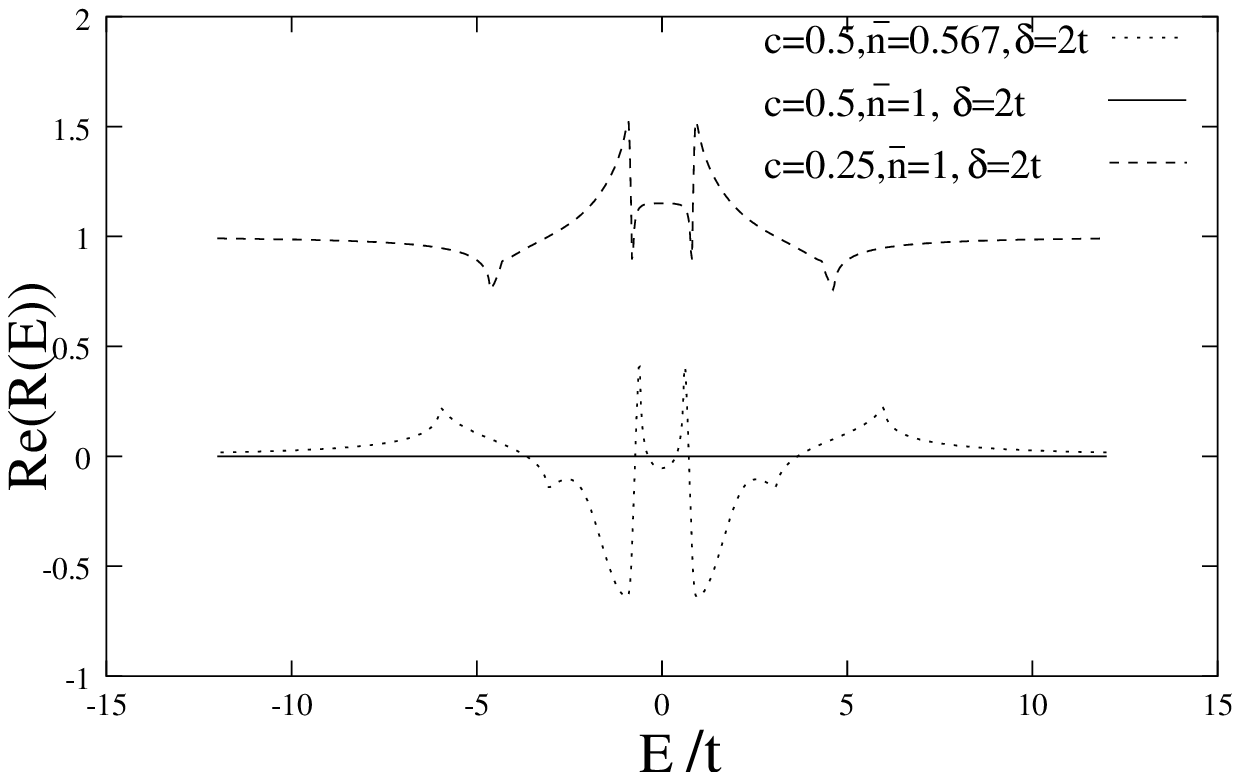,width=11.0cm,angle=0}}
\centerline{\epsfig{file=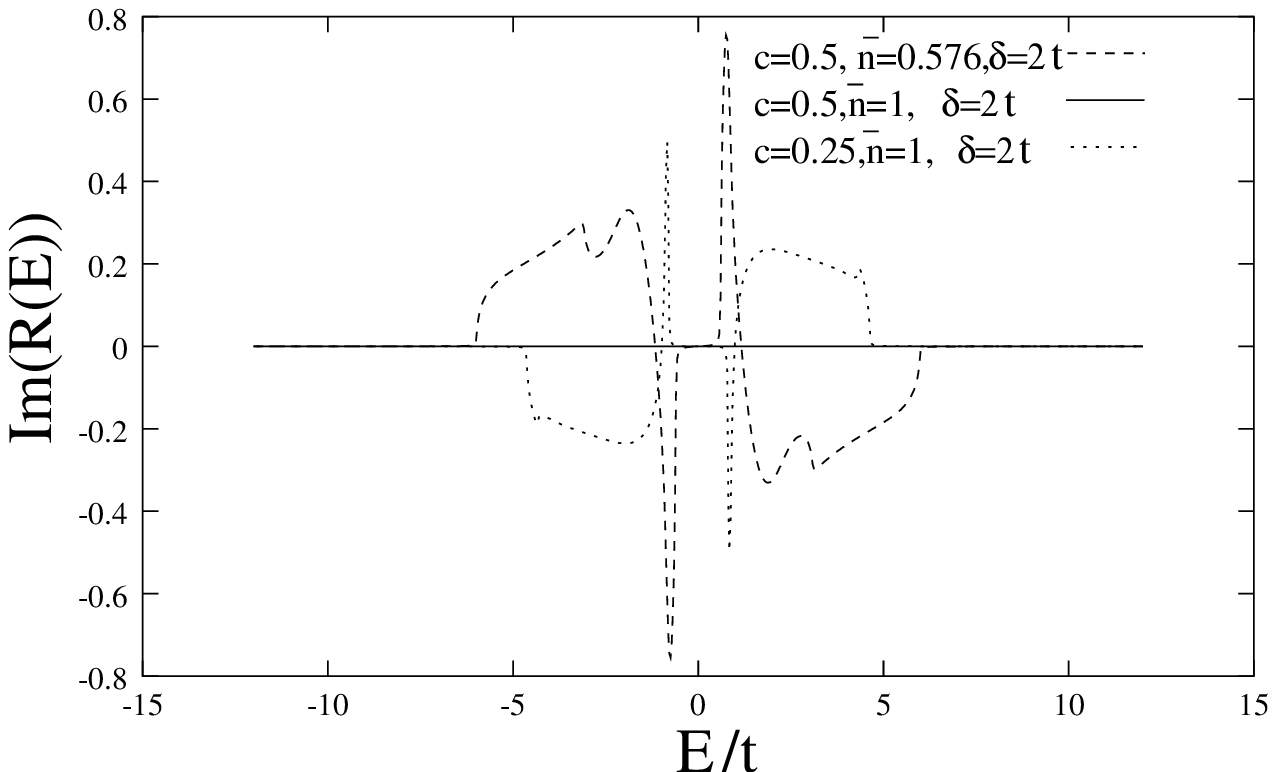,width=11.0cm,angle=0}}
\caption{ $R(E)$ is the analytical continuation of $\Re\Sigma_{11}(\imath\omega_{n})$ to the real axis. Here we show the real and imaginary part of $R(E)$, $\Re(R(E))$, $\Im(R(E))$ for three different cases: the cases $c=0.5$, $\bar{n}=0.576$, $\delta=2t$ and $c=0.25$, $\bar{n}=1$, $\delta=2t$ are
particle-hole asymmetric, and the case $c=0.5$, $\bar{n}=1$, $\delta=2t$ is particle-hole symmetric. The temperature here is 
 $T=0.008625t$. Clearly only in the particle-hole symmetric case is $R(E)=0$.
\label{figure6}}
\end{figure}.

\vfill
\begin{figure}
\centerline{\epsfig{file=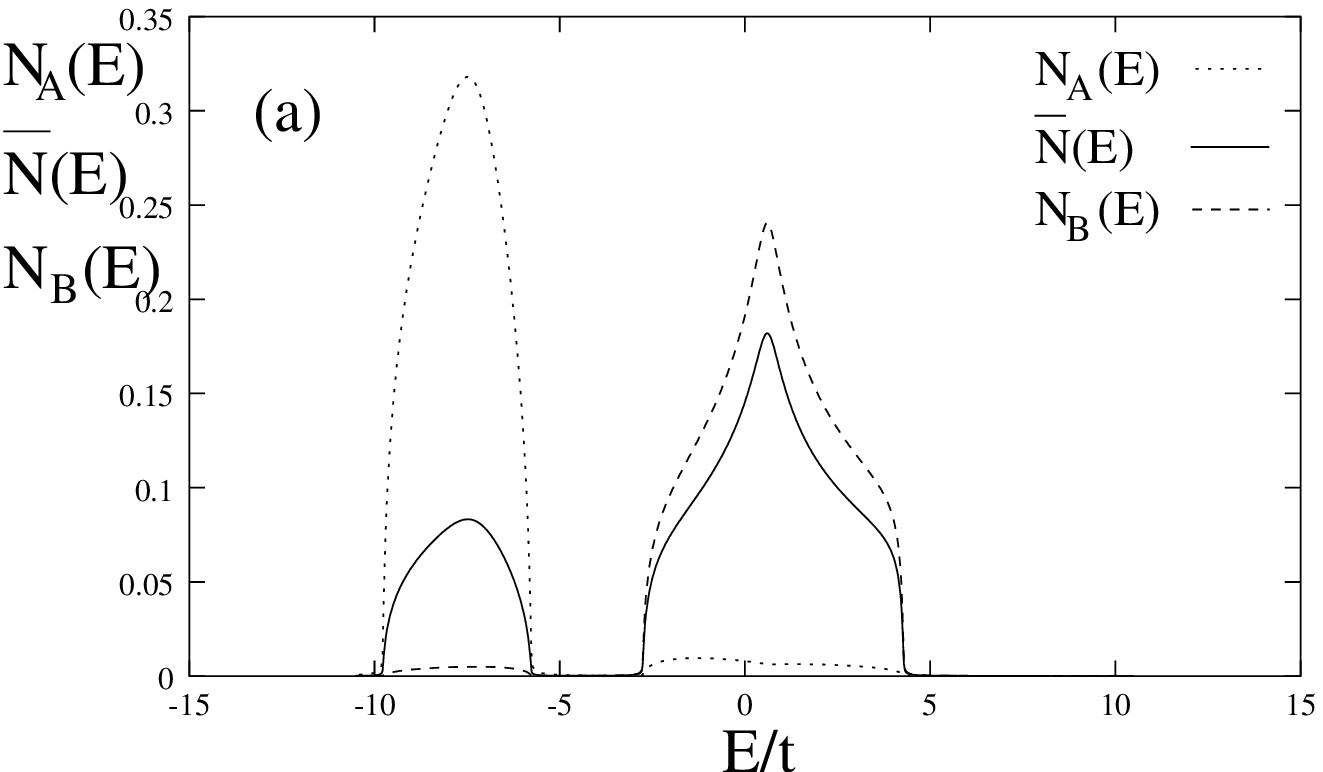,width=10.0cm,angle=0}}
\centerline{\epsfig{file=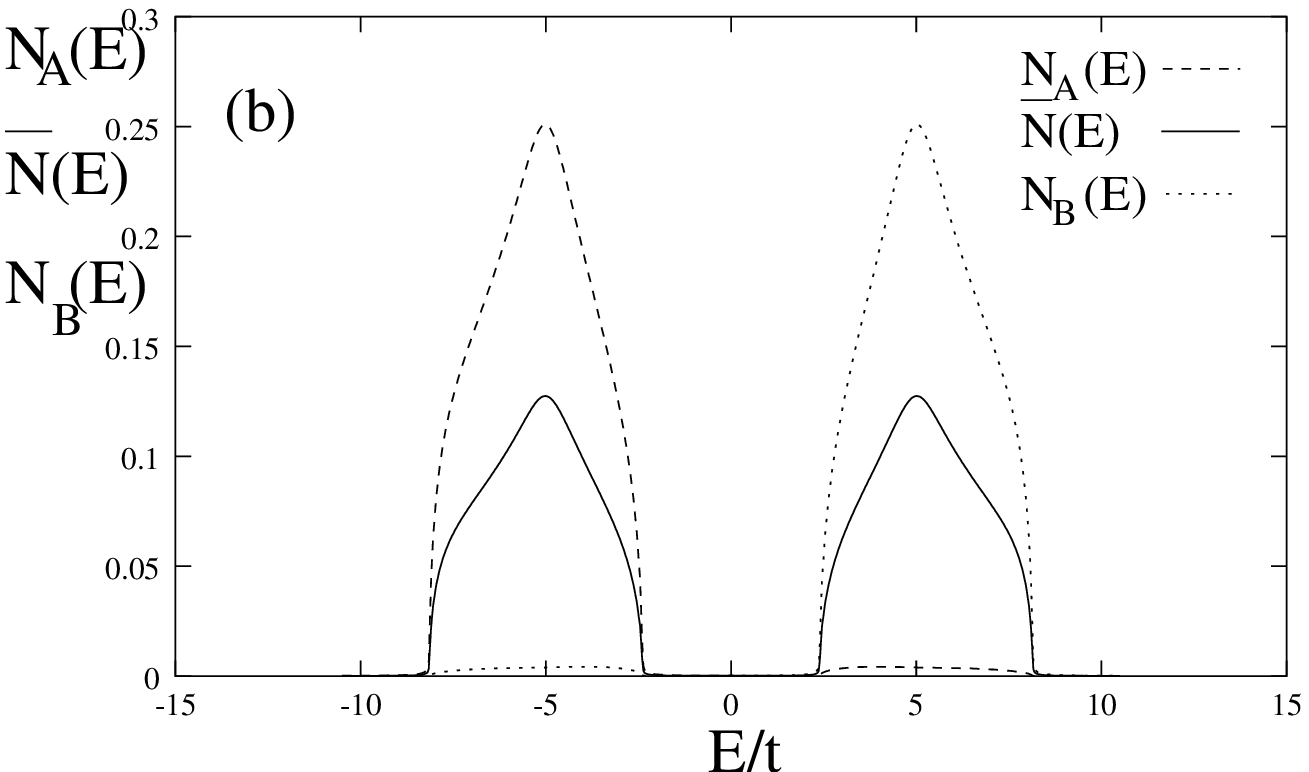,width=10.0cm,angle=0}}
\centerline{\epsfig{file=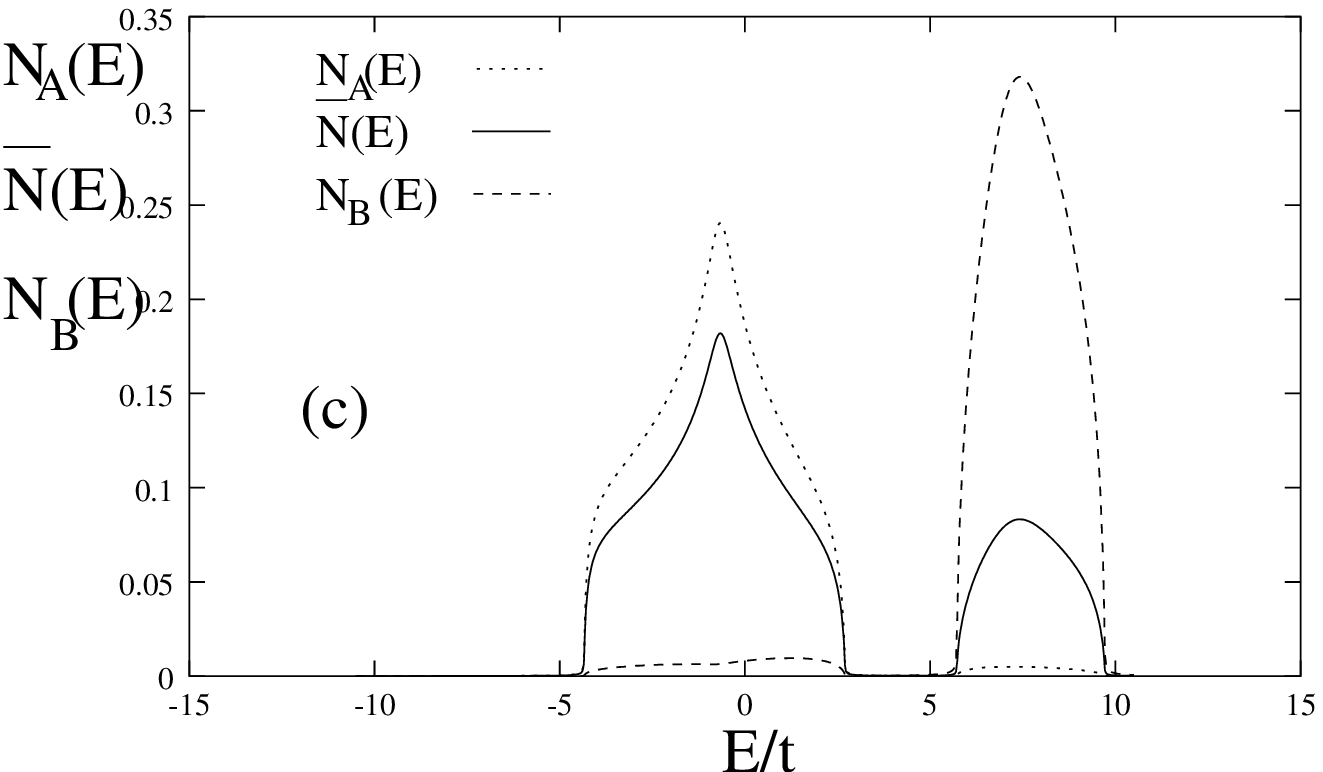,width=10.0cm,angle=0}}        
\caption{ Band splitting of the normal system at half band
filling $\bar{n}=1$ and $T=0.008625t$ in the three cases: (a) $c=0.25$, $\delta=8t$. In this case conduction is in the B band. (b) $c=0.5$, $\delta=10.1t$ in this case Fermi energy lies outside of both bands, therefore the state is an insulator. (c) $c=0.75$, $\delta=8t$. In this case the Fermi energy lies in the A band, so hopping occurs primarily in this band. 
\label{figure7}}
\end{figure}.

\vfill
\begin{figure}
\centerline{\epsfig{file=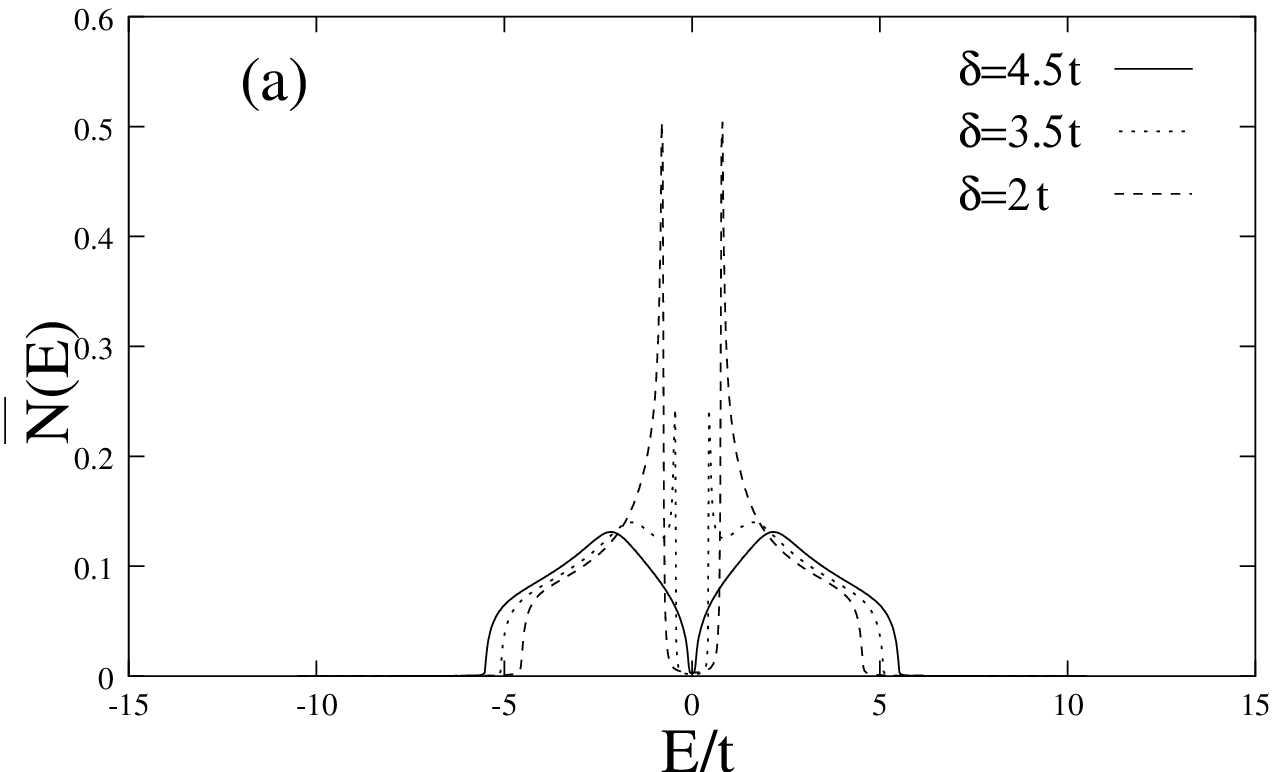,width=11.0cm,angle=0}}
\centerline{\epsfig{file=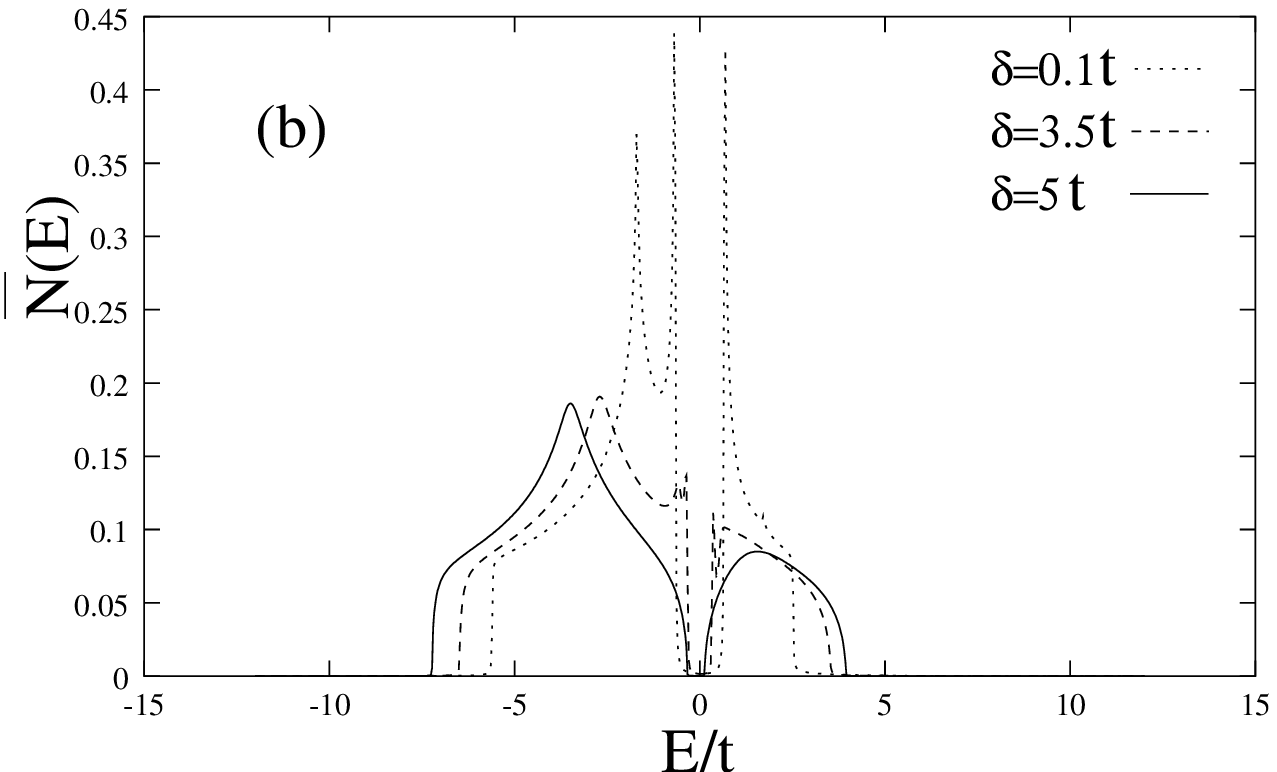,width=11.0cm,angle=0}}
\caption{ The superconductor-insulator phase transition in the strong scattering limit for two different alloy concentrations and average band filling at $T=0.008625t$. (a) A particle-hole symmetric density of
states at $\bar{n}=1$, $c=0.5$. (b) An asymmetric particle-hole density of
states  for $c=0.25$, $\bar{n}=1.5$. Note that in both cases the superconducting gap closes and is replaced by an insulating gap for large $\delta$.
\label{figure8}}
\end{figure}.

\vfill
\begin{figure}
\centerline{\epsfig{file=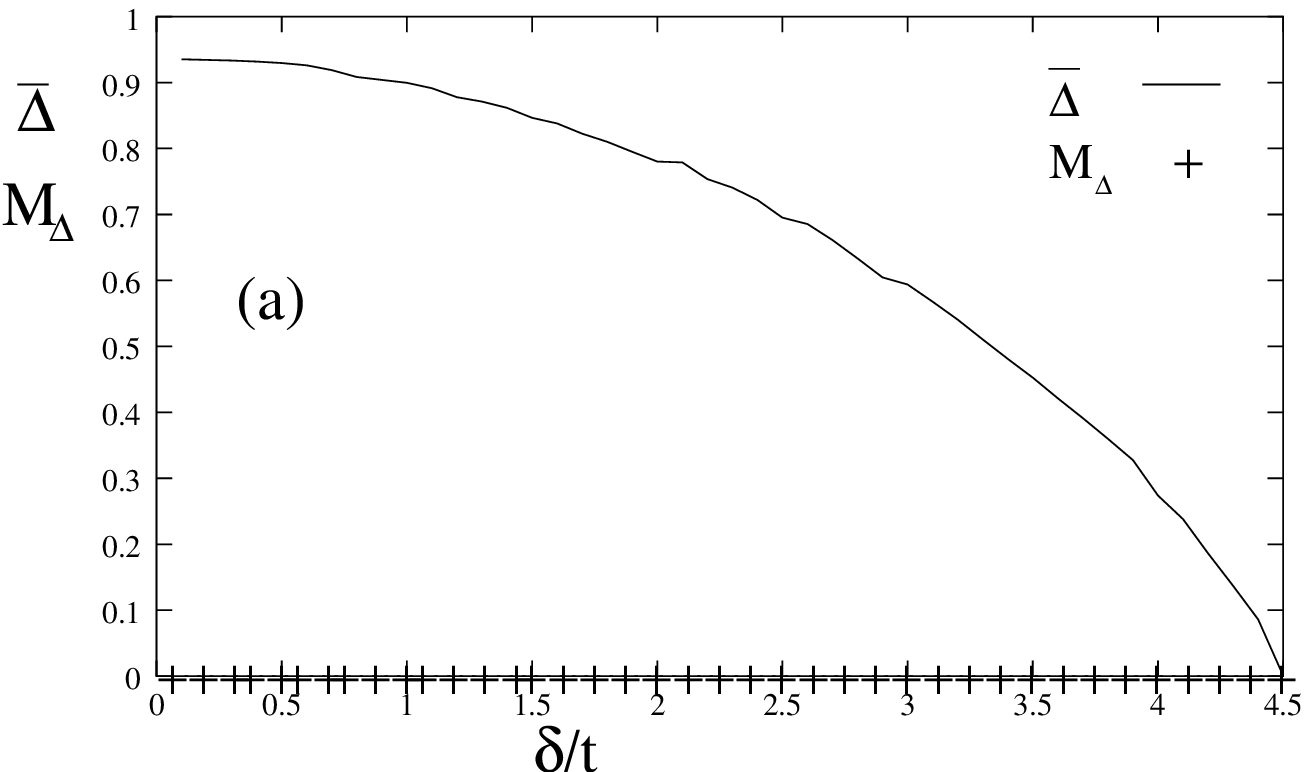,width=11.0cm,angle=0}}
\centerline{\epsfig{file=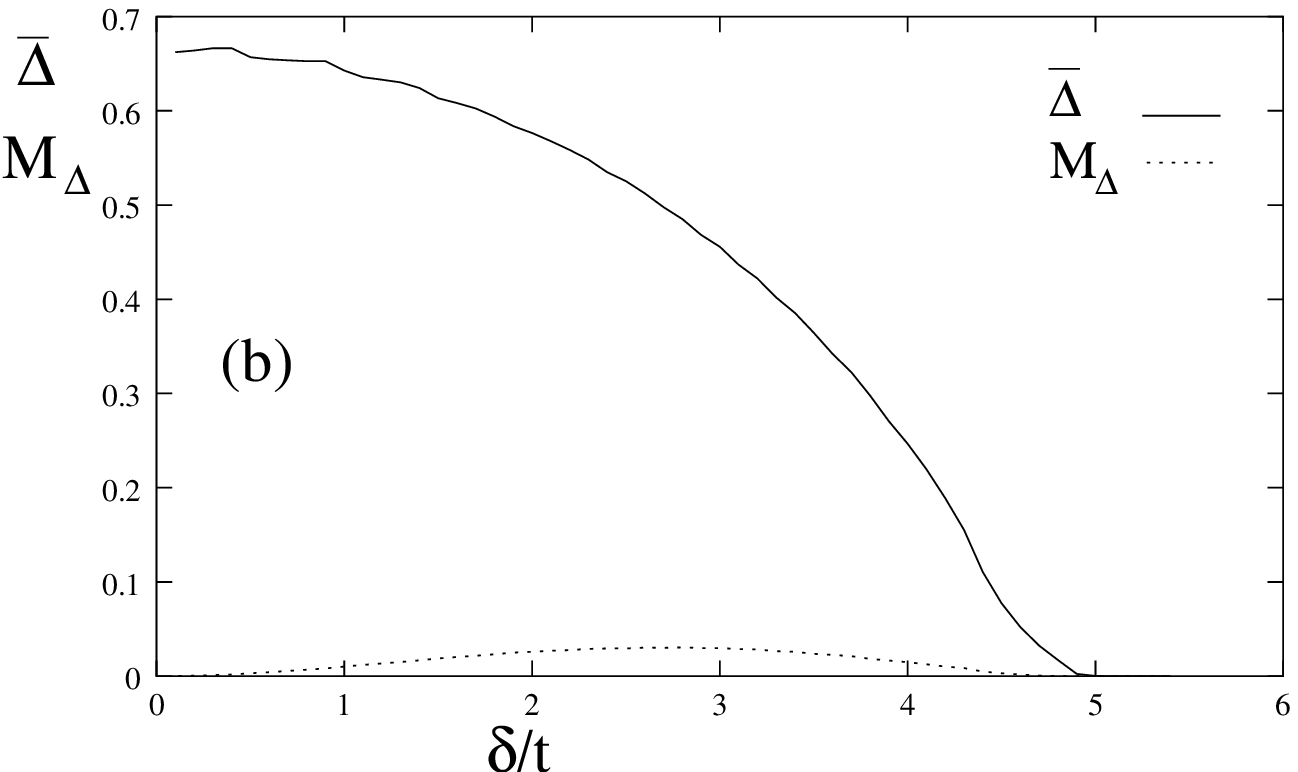,width=11.0cm,angle=0}}
\caption{ The fluctuation of $\Delta$, $M_{\Delta}=\langle
|\Delta^{2}_{i}|\rangle-{|\Delta^{c}}|^{2}$, and the average $\bar{\Delta}$ as a function of disorder strength. (a) 
The particle-hole symmetric case $c=0.5$ and $\bar{n}=1$, (b) A particle-hole asymmetric band $c=0.25$ and $\bar{n}=1.5$ at $T=0.008625t$. Note that the $\Delta$ fluctuation in the
particle-hole symmetric band is zero for all of disorder strengths.
\label{figure9}}
\end{figure}.

\vfill
\begin{figure}
\centerline{\epsfig{file=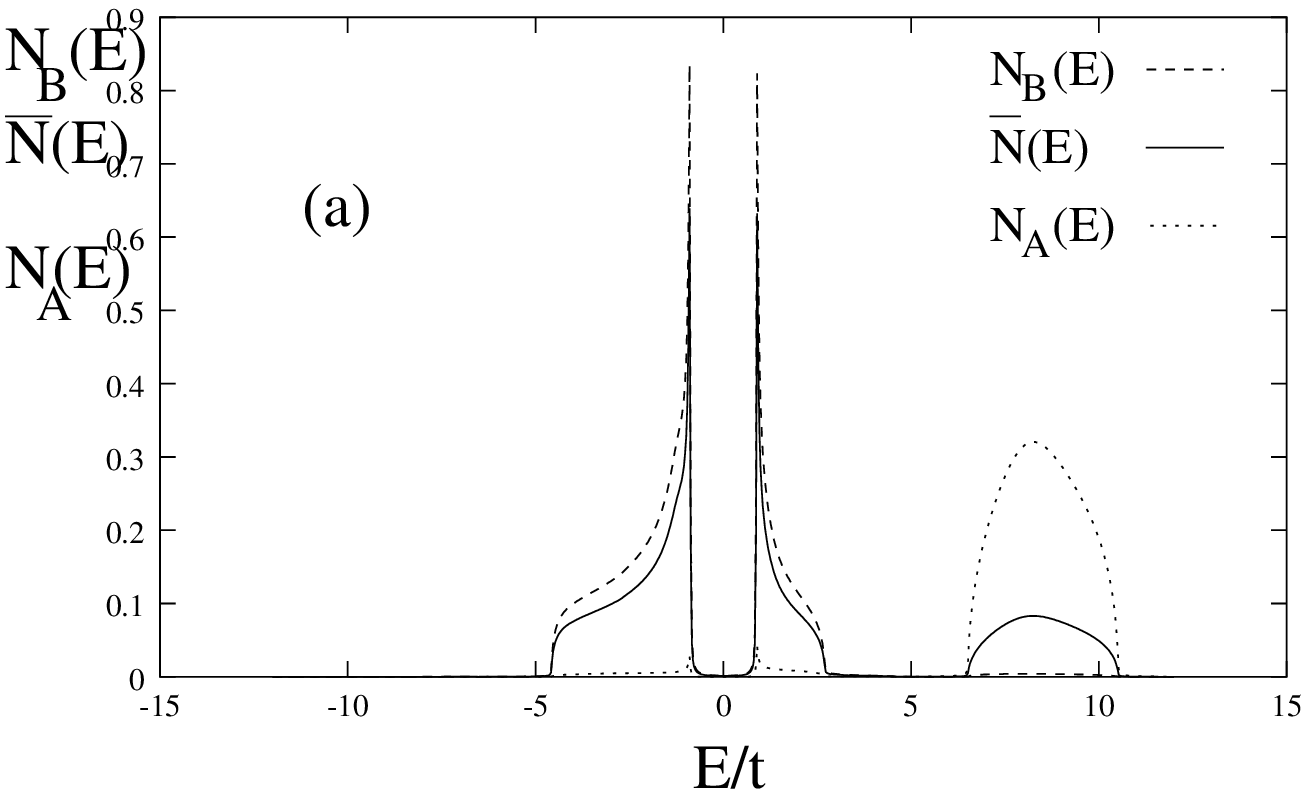,width=11.0cm,angle=0}}
\centerline{\epsfig{file=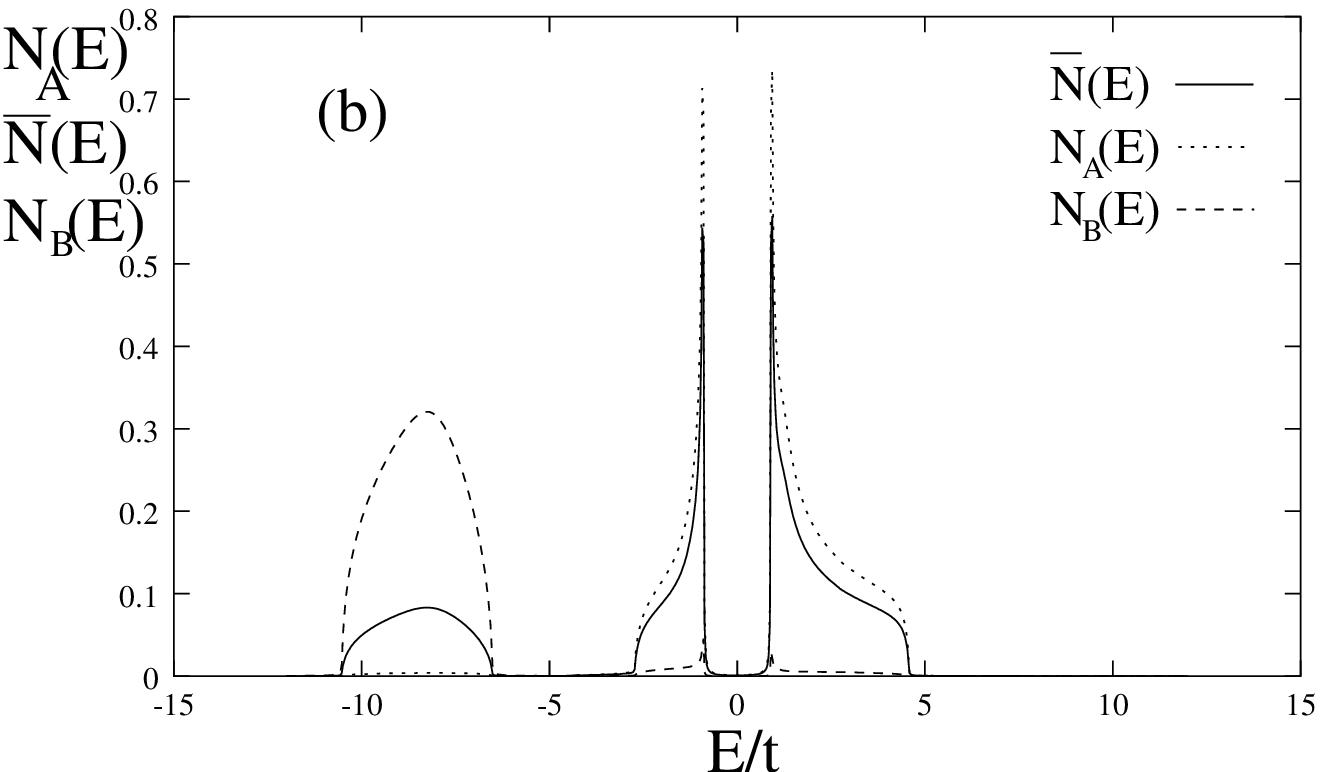,width=11.0cm,angle=0}}  
\caption{ Density of states at A and B sites $N_{A}$, $N_{B}$ and the average $\bar{N}$ at half band
filling, $\bar{n}=1$ and $T=0.008625t$, for the two cases:
(a) $c=0.25$, $\delta=9t$ and (b) $c=0.75$, $\delta=9t$. In the (a) case, A band is an empty normal band and superconductivity is only in the B band. In the case (b), the B band is a normal doubly occupied band and A band is a partialy occupied superconducting band.
\label{figure10}}
\end{figure}.

\vfill
\begin{figure}
\centerline{\epsfig{file=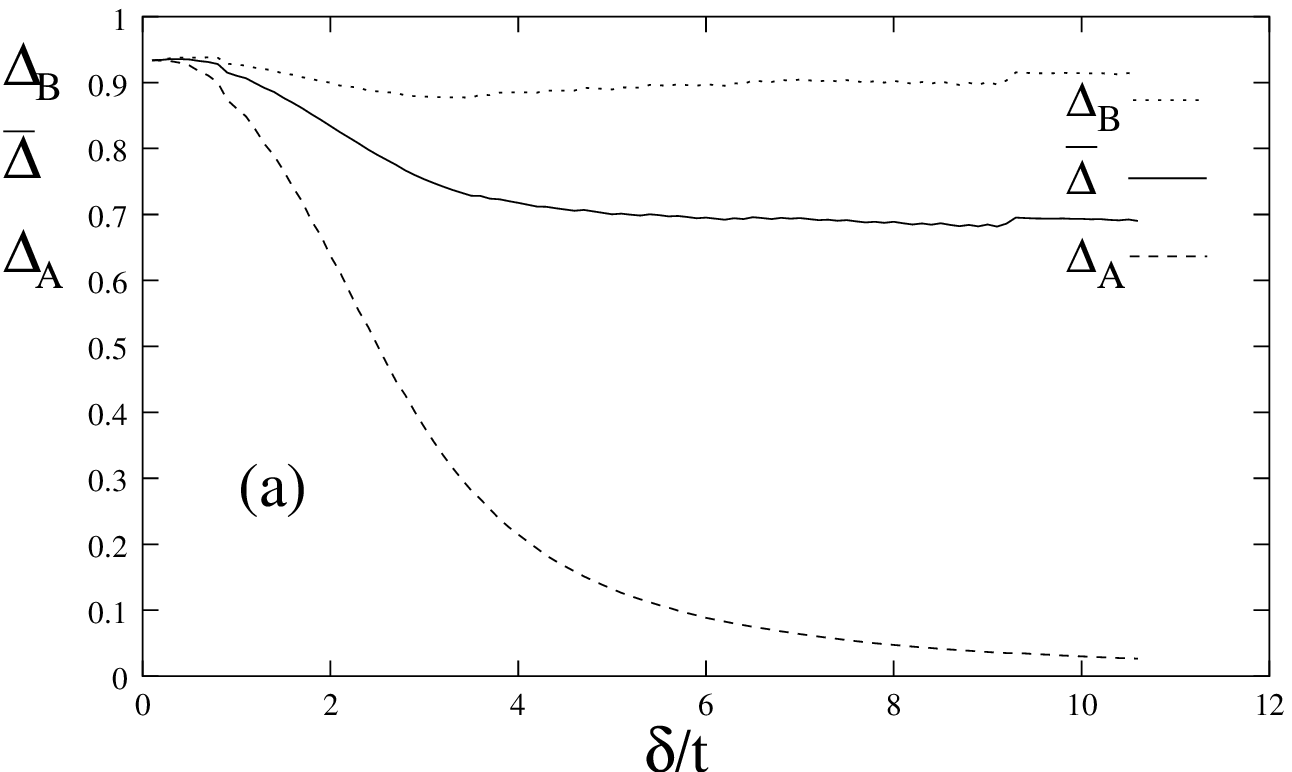,width=11.0cm,angle=0}}
\centerline{\epsfig{file=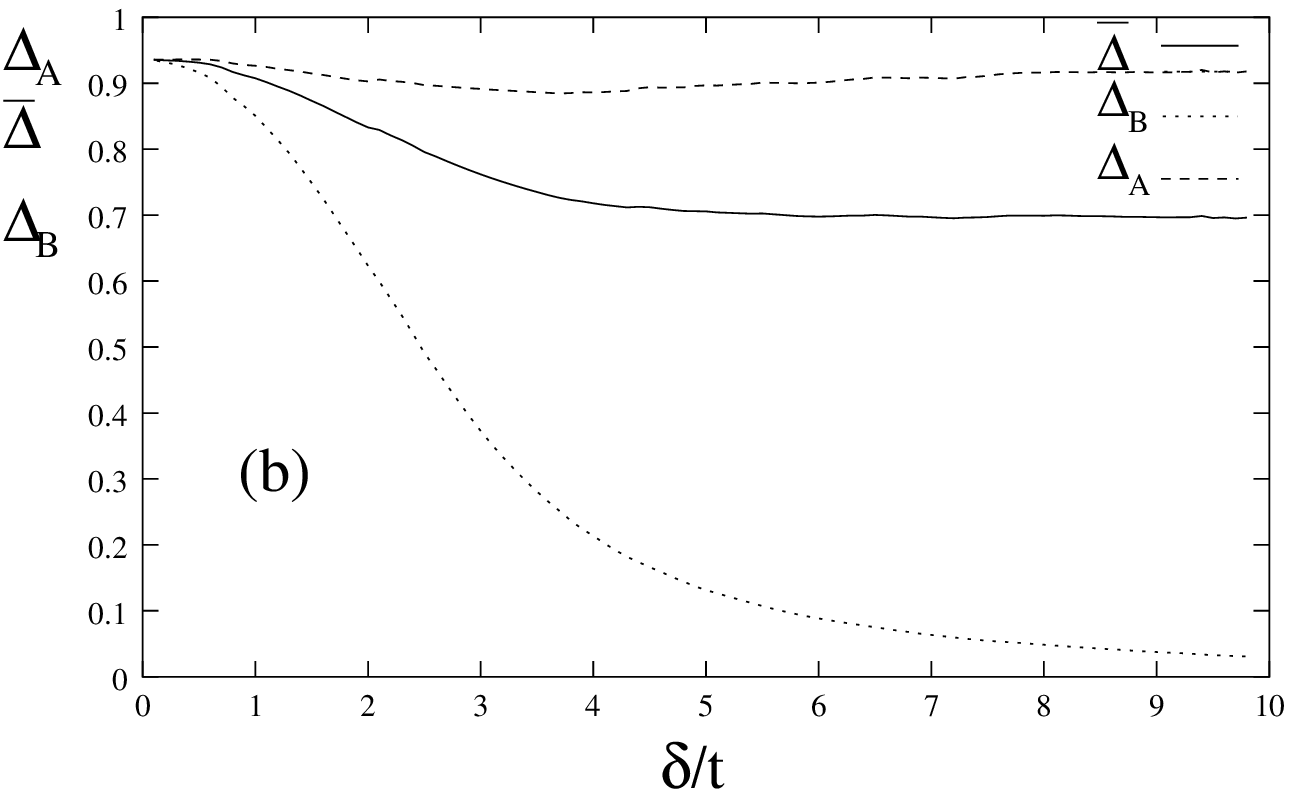,width=11.0cm,angle=0}}
\caption{ Order parameters $\Delta_{A}$, $\Delta_{B}$ and $\bar{\Delta}$ 
at half band filling $\bar{n}=1$ in terms of disorder strength for the  cases
 of: (a) $c=0.25$ and (b) $c=0.75$. For the case (a) and very strong scattering the order parameter on the A sites goes to zero. In case (b) the order parameter of the B sites goes to zero while that of the A sites goes to a constant.
\label{figure11}}
\end{figure}.

\vfill
\begin{figure}
\centerline{\epsfig{file=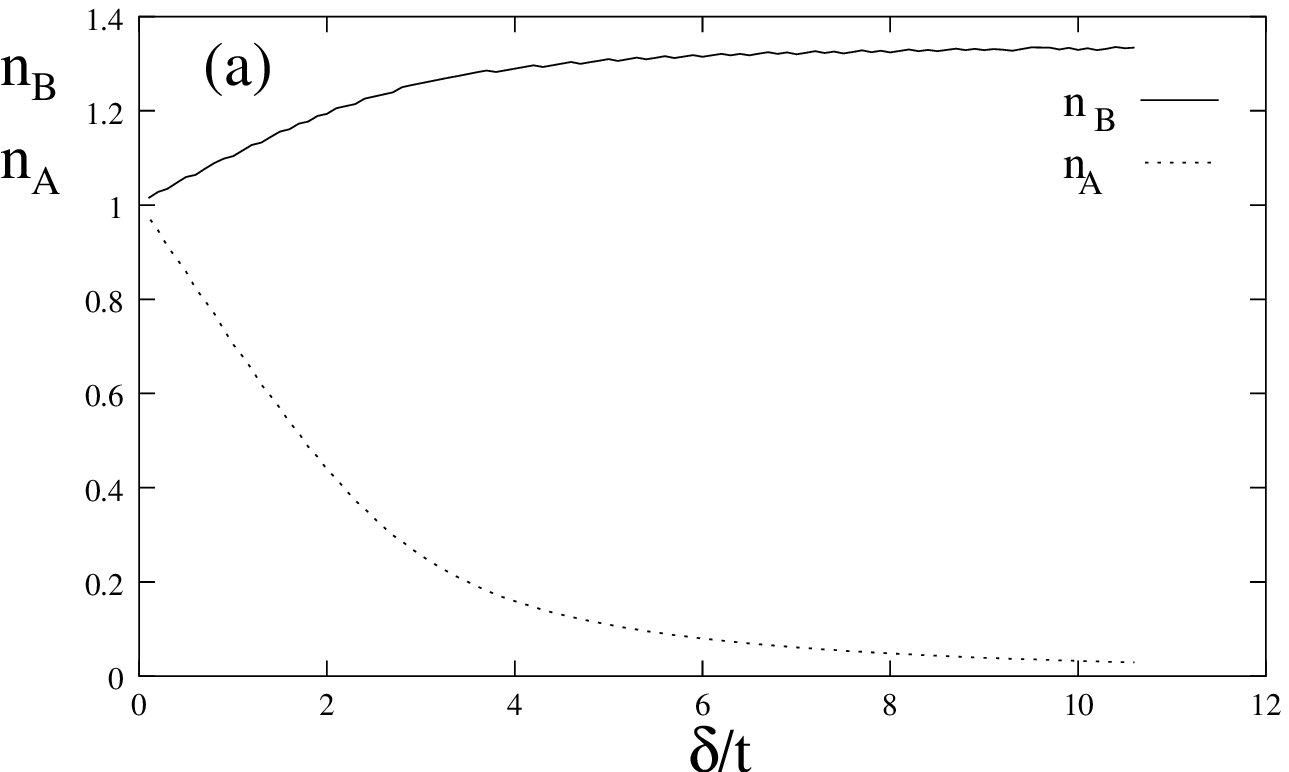,width=11.0cm,angle=0}}
\centerline{\epsfig{file=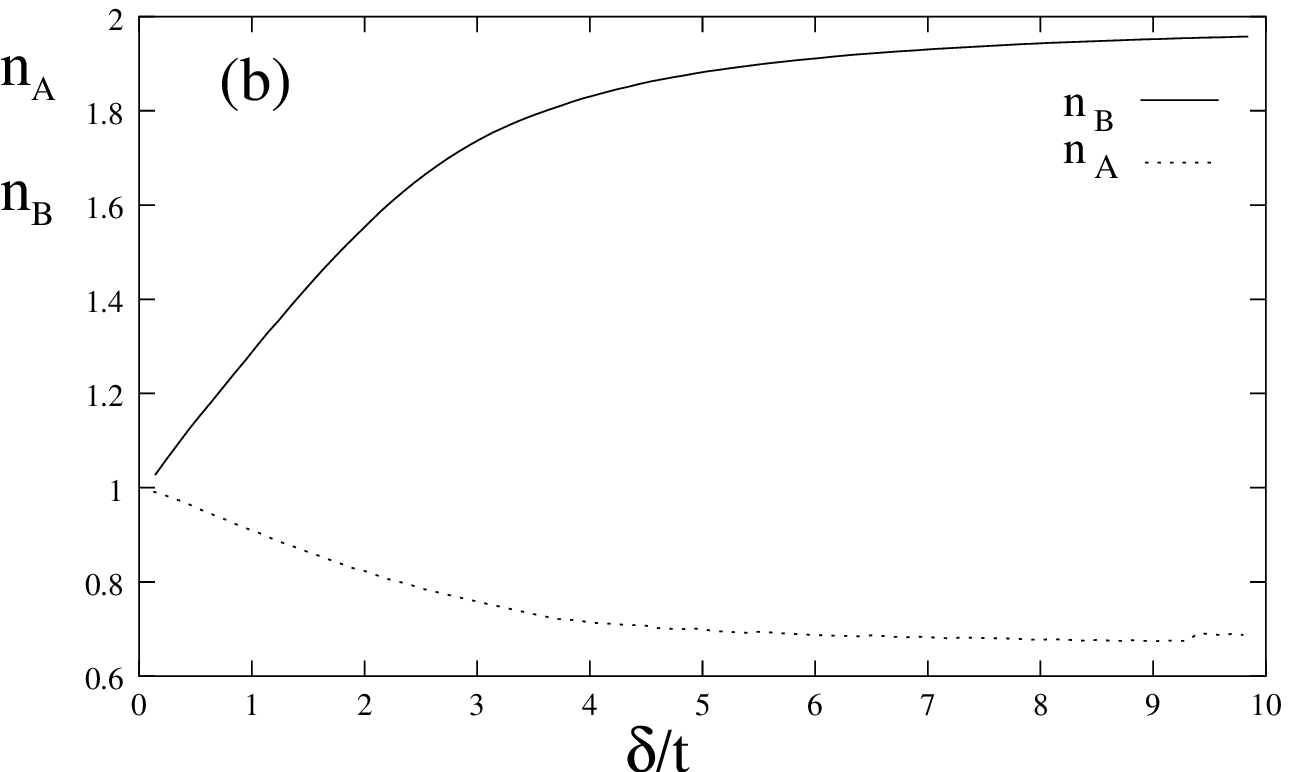,width=11.0cm,angle=0}}
\caption{ Charge densities $n_{A}$ and $n_{B}$ at half band
filling $\bar{n}=1$ and $T=0.008625t$ in
terms of the  disorder strength for the  case
 of: (a) $c=0.25$, (b) $c=0.75$. In the case (a) all of electons are in the B sites and A sites are empty in the strong disorder limit. The band filling of the B sites is $n_{B}=\frac{4}{3}$. But in case (b) the B band becomes fully occupied while the A band remains partially occupied.  
\label{figure12}}
\end{figure}.

\vfill
\begin{figure}
\centerline{\epsfig{file=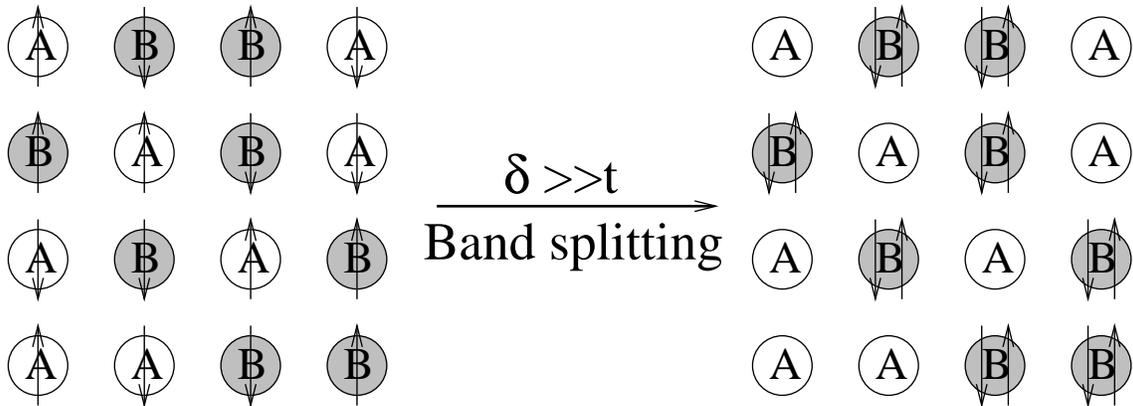,width=15.0cm,angle=0}}
\caption{ The physical mechanism of the band splitting in the particle-hole symmetric case of $c=0.5$, $\bar{n}=1$ for $T=0.008625t$. The band splitting happens at approximately $4.5t$. In this case all of the A sites become empty while the B sites become doubly occupied. Clearly this represents a band insulator.
\label{figure13}}
\end{figure}.

\end{document}